\documentclass[preprint,12pt,psfig]{aastex}

\newcommand{\be}{\begin{equation}}
\newcommand{\ee}{\end{equation}}

\newcommand{\ba}{\begin{eqnarray}}
\newcommand{\ea}{\end{eqnarray}}
\newcommand{\om}{\omega}

\newcommand{\Alfven}{ Alfv\'{e}n }

\newcommand\etal{\textit{et al.\ }}
\newcommand\eg{\textit{e.g.\ }}
\newcommand\cf{\textit{cf.\ }}

\begin{document}

\title{Magnetar giant flares and afterglows as relativistic magnetized 
explosions}
\author{Maxim Lyutikov}
\affil{
University of British Columbia, 6224 Agricultural Road,
Vancouver, BC, V6T 1Z1, Canada
 \\ and \\
 Department of Physics and  Astronomy, University of Rochester,
 Bausch and  Lomb Hall,
 P.O. Box 270171,
 600 Wilson Boulevard,
 Rochester, NY 14627-0171 }
 \email{lyutikov@phas.ubc.ca}


\date{Received   / Accepted  }

\begin{abstract}
We propose that giant flares on Soft Gamma-Ray Repeaters
produce relativistic, strongly  magnetized, weakly baryon loaded
magnetic clouds, somewhat analogous to
 solar
  coronal mass ejection (CME) events.
Flares are 
 driven by  unwinding  of internal non-potential magnetic fields which leads to 
 slow build-up of magnetic energy  outside of the neutron star. For large magnetospheric currents,
 corresponding to 
 a large twist of external  magnetic field,
 magnetosphere  becomes dynamically unstable on \Alfven crossing times scale
 of inner magnetosphere, $t_A \sim R_{NS}/c \sim 30 \mu$sec.
 The dynamic instability leads to formation 
 of dissipative  current sheets through development of tearing mode (Lyutikov 2003).
Released magnetic energy results in  formation of  a strongly magnetized, pair-loaded,
quasi-spherically
expanding flux rope, topologically connected by magnetic field to the neutron star during the prompt flare emission.
Expansion reaches large Lorentz factors, $\Gamma \sim 10-20$ at distances
$r\sim 1-2 \times 10^{7}$ cm, where lepto-photonic load is lost.
 Beyond this radius plasma is strongly dominated by magnetic field, though some baryon loading, with
 $M \ll E/c^2$, 
 by ablated neutron star material may occur. 
Magnetic stresses of the tied flux rope lead to late collimation of the expansion, on time
scales longer than giant flare duration.
 Relativistic
bulk motion of the expanding magnetic cloud, directed at an angle $\theta \sim 135^\circ$ to the line of 
sight 
(away from the observer), 
results in a strongly non-spherical forward shock with  observed non-relativistic  apparent
expansion and bulk motion velocities $\beta_{app} \sim \cot \theta/2 \sim 0.4 $ 
at times of first radio observations approximately one week after the burst.
Interaction with  a shell of wind-shocked ISM and then with the unshocked 
ISM leads to deceleration to non-relativistic velocities approximately one month after the 
flare. 
\end{abstract}

\section{Introduction}

Magnetar emission \citep[see,\eg,][for review]{wt04} is powered by dissipation
of  a  non-potential (current-carrying) magnetic field  \citep{tlk02}.
Dynamo mechanism operating at birth of neutron stars 
creates a tangled magnetic field inside a neutron star, which is prevented from unwinding by rigidity of 
the crust. 
Current-carrying plasma exerts Lorentz force on the crust, 
which is mostly balanced by lattice strain. 
For strong enough magnetic fields, Lorentz force may induce a  stress that exceeds the critical stress
of the lattice.
This leads to crustal motion, which should occur along equipotential surfaces:
crust will be rotating.
 Crustal rotation  and associated twist of magnetic field lead to expulsion of the electric  current from
inside of  neutron star into   magnetosphere. Dissipation of magnetospheric currents 
is  responsible for persistent emission \citep{tlk02}, while sudden reconfiguration of magnetic field may
produce flares \citep{l03}. Giant flare of SGR 1806-20 on December 27 2004 (we will refer to it as 
''the GF'') puts new constraints on the model that we discuss in this paper.

\section{
Where was energy  stored right before the flare? -- In the magnetosphere }

\subsection{Short rise time}

GFs are powered by dissipation of magnetic field energy.
One of the principal issues is where 
most of the magnetic energy has been stored prior to the GF: 
in the magnetosphere or in the neutron star crust.
These two possibilities are related to two models of giant flares of the SGRs.
First, a giant flare may result from a {\it  sudden }
untwisting of the internal magnetic
field \cite{td95,thompson01},  TD95 and TD01 below.
In this case, a large and quick (on time scale on a flare)  rotational displacement
of the crust leads to {\it increased  twisting  of  magnetospheric magnetic field lines}.  
Alternatively, {\it slow }  untwisting  of the internal magnetic field
leads  to gradual twisting of  magnetospheric  field lines, on time scales much longer than GF, until 
it reaches a dynamical stability threshold  due to increasing energy associated with
current-carrying magnetic field. Then
{\it sudden relaxation } of the  twist outside the
star and associated dissipation and magnetic topology change
lead to flares,  in  analogy with Solar flares and Coronal Mass Ejections (CMEs).
Note, that even in case of crustal storage of magnetic field energy before the flare  (TD95, 2001),
dissipation  also occurs in the magnetosphere, not in the crust.

The best test of these two alternatives is the  time scale for the
development of a 
flare. Since energy involved in the GF requires that a large fraction
of the magnetosphere is affected, a typical size of the active region is of the order of 
the neutron star radius. Sudden unwinding of the crust should occur  
 either on shear wave or \Alfven wave
 crossing time of the star  $t_{A, NS} \sim R_{NS}/V_{s,NS} \sim   R_{NS}/V_{A,,NS}\sim 0.2-0.5$ s 
 (TD95), 
 while magnetospheric instability may develop on time scales as short as 
  \Alfven
   crossing time of the inner magnetosphere, $t_{A, ms} \sim R_{NS}/c \sim 30 \mu $s \citep{l03}.
   Observations of the December 27  GF show very short rise time $\sim 0.25 $  msec
   \citep{palmer05}.
{\it Very short rise time of the GF
 points to magnetospheric origin of GF} 
(in a sense that right before the flare the energy to be released is stored in the magnetosphere).

\subsection{Pre- and Post-burst evolution of persistent emission}

XMM-Newton observation of SGR 1806-20  months before the GF
   shows an  increased activity,
 with persistent flux increasing  by a factor of two, spectrum hardening
  (photon power law index decreased from 2.2 to 1.5) and spindown increasing \citep{turolla},
 all in agreement with the
 prediction of  twisted magnetosphere model \citep{tlk02}, implying an 
 increasing twist 
 before the GF.
 In addition,
two months after the GF  
pulsed fraction and the spin-down rate
have significantly decreased and  the spectrum softened \citep{rea05}.
The same occurred in  SGR 1900$+$14 following
the August 27 GF \citep{woods01}.
All of these effects 
are consistent with 
increasing of the twist during the time leading to the GF and 
 decreased  twist of external magnetic fields after the  GF, brought about
by reconnection: (i)
in the reconnection  model the post-flare magnetosphere
 is expected to have  a simpler  structure, as the pre-flare network of currents
 has been largely dissipated; (ii) non-thermality of the
 spectrum is a measure of the current strength in the bulk of 
 magnetosphere \citep{tlk02,lf05} with  softer  spectra corresponding
 to a smaller twist; 
 (iii)  spin-down rate depends on the amount of
 current flowing through open field lines and is 
 smaller for a  smaller twist. Note, that since
 open field  lines occupy only a small
 fraction of magnetosphere,
 spin-down rate probes  current in 
 a relatively small region which
 should, on one hand,
 correlate with typical current in the magnetosphere 
 on long time scales, but on the other hand may show large deviations 
 on short time scales.

\subsection{Ejecta must carry  a lot of magnetic field}

Magnetospheric storage and  release of  energy  leads to the following consequences.
First,  it is hard to see how most of the energy released in the magnetosphere
can be spent on heating the surface of the neutron star and generating heavy ion-loaded outflows.
 Secondly, dissipation of magnetic field  cannot create magnetic field-free plasma:
 it is likely to be limited to equipartition fields
since at this point induced magnetic field of gyrating relativistic particles will create magnetic field
comparable to the initial field (at temperatures near $500 $ keV energy in
photons will be comparable to energy in electrons).
Thus, only approximately half of the magnetic field
is expected to be converted to particle and photon energy, not much more. 
Most of the dissipated energy will be later radiated away and/or spent on $pdV$ 
work during expansion.

Thus, magnetospheric release of energy, indicated by very short  rise time of the GF, 
decreasing persistent emission and softer post-flare spectrum, 
leads  to the conclusion that expanding plasma must be strongly magnetically dominated.
In this paper we examine consequences and consistency of the model based on these 
premises.

\section{Overview of the model}

Before discussing various details, let us present a short overview of the model, which qualitatively
resembles models of Solar flares 
with the difference that Sun supports an actively operating dynamo,
while in magnetars dynamo operated only during  birth of a neutron star.
The energy that will be released in a GF is initially (at times long before the flare)
 stored in  electric currents
flowing  inside  neutron star. 
These currents
are generated during birth of the neutron star
and  are slowly pushed out into magnetosphere, gated by {\it slow, plastic deformations}
of neutron star crust. This creates active magnetospheric regions, in analogy with Solar spots.
An active region consists of a sheared arcade of magnetic flux 
and surrounding non-potential magnetic structures.
As currents  (and with them magnetic  energy and helicity) 
are pushed outside the neutron star, magnetosphere adjusts
slowly to the changing boundary conditions. During this phase magnetic energy
is slowly  stored in the magnetosphere. As more  current is pushed outside,
magnetosphere  reaches a point of dynamical
instability beyond which stable equilibrium cannot be maintained.
Crossing the instabilities threshold leads to  changing 
of magnetic configuration on   \Alfven crossing time scale, formation of narrow current sheets, 
and onset of magnetic dissipation (this
  process is sometimes called magnetic detonation
\citep[\eg][]{Cowley}). 
This has two consequences. First, a large amount of magnetic energy
is converted into kinetic plasma  energy and photons. Secondly, dissipation allows
to change magnetic topology and leads to formation of an expanding 
magnetic loop that eventually break  away from the star. Initially, after onset of reconnection,
kinetic pressure of optically thick  pair plasma
and magnetic stresses  are comparable,
so that 
 expansion is  quasi-isotropic and reaches relativistic  Lorentz factors
 $\geq 10$, determined  either by
  baryon loading or, in case of very small  baryon loading,
  by the  amount of residual pairs. 
  
\begin{figure}
\plotone{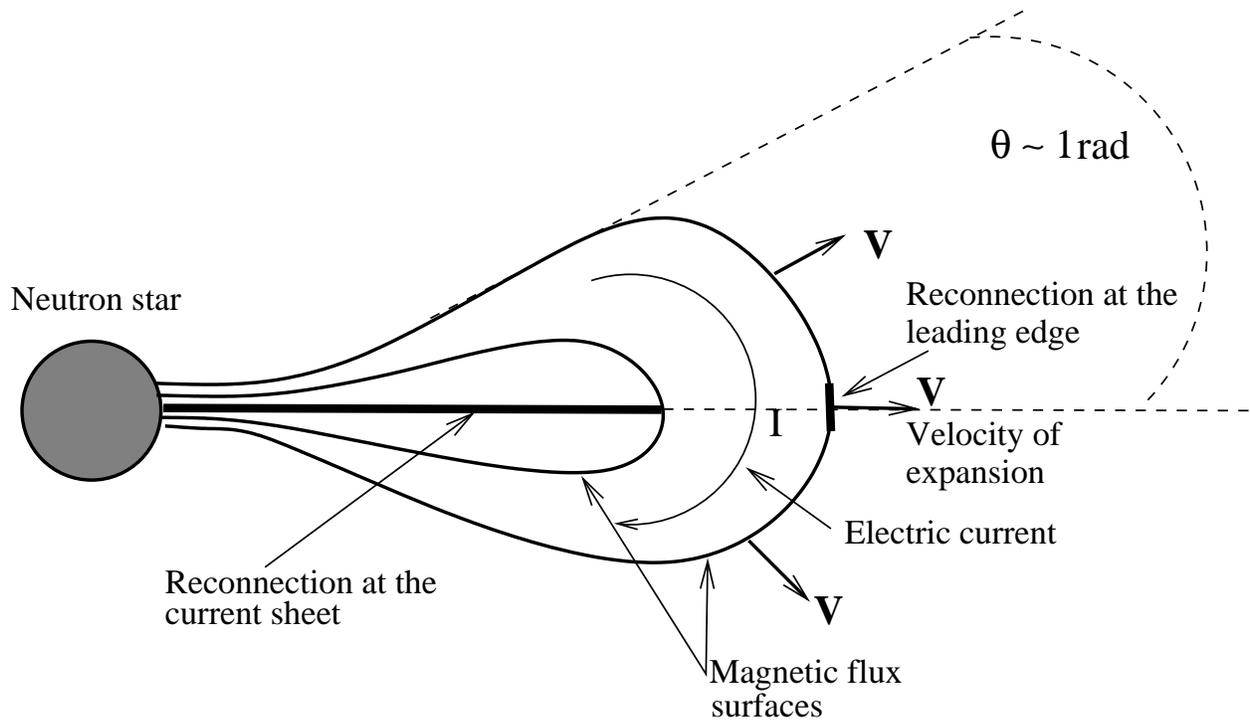}
\caption{ Cartoon of an expanding, collimated flux rope. 
Twisting of footpoints of a flux tube leads to electric current flow along the loop 
and results in its expansion. Beyond some twist angle
dynamic instability leads to formation of dissipative current sheets at the leading edge of the loop,
between the footpoints and, possibly in the bulk. 
Reconnection at the leading edge allows the flux tube to break out of the magnetosphere. 
Collimating effect of the tied footpoints later lead to broadly  collimated outflow.}
\label{fluxrope}
\end{figure}

  During the prompt phase of the GF, the expanding magnetic  
  {\it loop remains attached to the star},
 see Fig. \ref{fluxrope}. 
 Tying of the expanding loop to the star eventually leads to collimation
 of the explosion into a wide opening angle, of the order of one steradian.
 After losing the  pair load, the expanding cloud is dominated by magnetic field (magnetic cloud) and 
 eventually disconnects  form the neutron star, moving relativistically
 {\it away} from the observer at an
 angle $\sim 135^\circ$ degrees. This results in apparent subluminal 
 proper expansion and proper velocity. Eventually, energy of the magnetic cloud
 is transfered to the strongly anisotropic forward shock which produces
 the observed afterglow radio emission.

\section{Prompt and tail  emission of the GF }

\subsection{Twisting of external magnetic field and GF precursors}
\label{precursors}

 The central point of our suggestion is that
 winding-up of external magnetic field proceeds on long time scale, much longer than
  GF. 
The winding can occur on time scale of approximately 
  two months before the GF, indicated by increased activity of the source  \citep{turolla}.
 Alternatively, onset of fairly rapid plastic deformation of the crust may be related to
 weak emission events (precursors) that seem  to  precede GFs.
 [In case of GF of SGR 1900+14 a  precursor was seen $\sim 0.4 $ seconds before the main burst
 \citep{feroci01} while in case of   SGR 1806
 a relatively  powerful event occurs approximately hundred  fourty seconds
 before the main burst \citep{palmer05}.]
 During the quiescent period between the precursor and
 GF, a patch of a crust is continuously rotated by the Lorentz force,
   balanced both by elastic and  viscous stresses in the crust.
 These two possibilities are not mutually exclusive: slow evolution on time scale of months
 may be followed by a relatively fast twisting over hundreds of
 seconds before the GF.

Consider a crustal plate of size $R$ rotating under the influence of Lorentz force, balanced by 
viscous stresses at the base of the crust.
The 
dissipated power  is \citep{LLIV}
\be
L_{visc} \sim {4 \sqrt{2} \pi^{7/2}  \sqrt{\nu} R^4 \rho \over T_{rot}^{5/2}} =
1.3 \times 10^{37} {\rm erg s}^{-1} 
\left( { R \over 10 {\rm km} } \right)^{4} \,
\left( {\rho \over 10^{14} {\rm g cm}^{-3} }  \right)
\left( {\Delta \phi \over 2 \pi} \right)^{5/2} 
\ee
where $\nu \sim 10^4 \left( {\rho \over 10^{14} {\rm g cm}^{-3} }  \right)^{5/4} $ 
is the viscosity of neutron star \citep{cutler}, $\rho$ is density at the base of the crust, 
$T_{rot}=140 \left( {\Delta \phi \over 2 \pi} \right) $
s is the rotation period of the plate, taking into account 
that instability occurs after rotation of $\Delta \phi $ radians. 
Total viscously dissipated energy is $E_{vis} \sim L_{visc} T \sim 2 \times  10^{39} 
$ erg.
[For  SGR 1900+14 with $T_{rot} =0.4$ sec  $L_{visc} \sim 3 \times 10^{43} erg {\rm s}^{-1}$ and
$ E_{vis} \sim 10^{44} $ erg,
but a very short $T_{rot}$ may indicate that twist was near critical
before onset of rotation,
$\Delta \phi \ll 1$.]
Since this energy is released deep in the crust, where thermal
diffusion time  to the surface is much longer than $T_{rot}$,
 most of the heat is absorbed by the core \citep[\cf][]{lyubar}
and  does not show as increased 
persistent emission between the precursor and the main flare. 

During plastic creep, 
elastic strain is much larger than  plastic strain.
The  difference between the two
can be expressed in terms of how much magnetic field exceeds field at the critical strain:
\be 
\Delta B \sim 2^{9/4} \pi^{3/4} \sqrt{R \rho} \nu^{1/4} T_{rot}^{-3/4} =
2.7 \times 10^{10} {\rm G} \left( { R \over 10 {\rm km} } \right)^{1/2} \,
\left( {\rho \over 10^{14} {\rm g cm}^{-3} }  \right)^{1/2}\, 
\left( {\Delta \phi \over 2 \pi} \right)^{3/4}
\ee
($\Delta B = 2.2 \times 10^{12} $ G for  SGR 1900+14).
Thus, during plastic creep 
only a small fraction of magnetic energy is dissipated in the crust, most of it 
is  pumped outside of the
star.

Note that in case of the Sun,
TRACE satellite  has
detected rotation of  sunspots associated  with largest (X-class) flares  days  before
the flare \citep{Mewaldt}.
In addition,  Solar CMEs also starts before the accompanying X-ray flare: there is a quiet growth period  of approximately 30 minutes
 before formation of the dissipative current sheet \citep{zirin88}.

\subsection{During the giant flare expansion must  be relativistic }
\label{relativistic}

Relativistic expansion at the time of the GF follows from the 
conventional compactness argument. For luminosity 
$L \sim  10^{47} $ erg/s and variability time scale
$\sim 1 $msec,
optical depth to pair-production is \citep[see also][]{nakar05}
\be
\tau_{\gamma-\gamma} \sim {L \sigma_T \over 4 \pi m c^3 R_{NS} }
\sim 2 \times 10^{11}
\ee
 If plasma were to remain 
non-relativistic, photon diffusion times would be long, inconsistent with short 
observed variability time scale. 
This estimate immediately  excludes large baryon loading: $M$ must be $\ll E /c^2$.  
This  has always been a standard view of giant flares, \citep{feroci01,thompson01}.
[A possibility that initial $\gamma$-ray spike was produced in a  relativistic outflow
 without  dynamically important
 magnetic field and did
not contribute  significant amount of energy to afterglow
is hard to reconcile with magnetospheric origin.]

At early stages dissipation of magnetic field creates an optically dense lepto-photonic plasma
emerged in magnetic field
with $T \sim (L / 4 \pi R_{NS}^2 \sigma_{ST} )^{1/4} \sim 300 $ keV 
\citep[see also][]{nakar05}.
Qualitatively, quasi-spherical expansion of a strongly magnetized pair bubble
resemble the unmagnetized case, but there are important differences in the asymptotic
dynamics, 
outlined in  Appendix \ref{warm}. Initially,
plasma  expands  with bulk Lorentz factor increasing approximately linearly 
with radius $\Gamma \propto r$, while rest temperature decreases $T  \sim 1/r$. 
After reaching $ T _\pm \sim 20 $ keV at $ r_\pm \sim   1.5 \times  10^7$ cm (at which point
$\Gamma \sim 15$), plasma
becomes optically thin. Observed emission is  thermal with 
$T_{obs} =  \Gamma T _\pm  \sim 300 $ keV.

The main
implication of relativistic expansion is that outflow is {\it not} heavily loaded 
with baryons. In \S \ref{Basicafter} we will show that this picture is consistent with
 {\it observed } non-relativistic expansion velocities of the afterglow.

\subsection{ Initial millisecond spike is nearly isotropic}
\label{Initial}

If initial spike were  strongly anisotropic, with luminosities inside and outside
 some emission cone different by orders of magnitude, 
then we would have observed many more tails without initial spikes.
Not  a single tail without a spike has been seen.
Tails 
 are obviously only weakly
anisotropic, emitted by a trapped fireball (TD95), while their 
intensity is well above threshold of detectors.

Since initial $\gamma$-ray emission is nearly isotropic, it is unlikely
to be produced by a strongly  jetted outflow  \citep[contrary to ][]{Yamazaki}.
It is still feasible that the  initial spike is weakly anisotropic, with  radiation
intensity and
Lorentz factors  changing  by some factor $   \leq 2$ depending on direction.
[Note, that tail emission in all cases was of the same order, while the energies of the
initial spikes 
were vastly different. This fact is, on one hand, consistent with  some structured 
jetted emission  of the initial spike, so that  all bursts are the same 
but initial spike is viewed from different angles \citep{Yamazaki}, but on the other hand it
 contradicts the
fact that afterglow emission in case of the GF was several orders brighter, 
arguing in favor of larger total energetics. Constant tail emission may be explained as 
a limiting effect of magnetar magnetic field:  above some threshold  
flare energy,  the amount of trapped plasma depends only on the 
strength of confining 
poloidal magnetic field  and not on the amount of the released twist.]

\subsection{Time scales of the GF}
\label{T}

There are several time scales in the model: first is the  slow initial twist of external magnetic field
\S \ref{precursors}. Secondly, there  are several time scales associated with the GF itself:
(i) sub-millisecond initial rise  $.25 $ msec (ii) $\sim 5$ msec  rise to the main peak, (iii)
hundreds millisecond total duration of the spike, (iv) tens of seconds tail emission.
Finally, there is afterglow time-scale from one week to $\sim$ one month, the latter   we 
identify with non-relativistic transition of the 
ISM blast wave.

Let us discuss the time scales of the GF itself.
Primarily, we associate the shortest time scale observed in the burst $\sim 0.25$ msec
with the \Alfven  crossing time of the inner magnetosphere.
It reflects the dynamical evolution of the magnetosphere after it has crossed
stability threshold. Thus, the very first photons are emitted while
plasma is still not expanding relativistically.

 Secondly, since  emission requires dissipation of energy,
dissipative  time scales  become important as well. In magnetar {magnetosphere}s 
 development of dissipative tearing instability  occurs on time scale intermediate
 between \Alfven times scale and resistive time scale: 
 $t_{tearing} \sim \sqrt{ \eta c /R_{NS}}$ where $\eta$ is plasma resistivity. 
 If resistivity is related to plasma skin depth, $\eta \sim c^2 /\om_p$, 
 where $\om_p$ is plasma frequency, the growth rate 
 of tearing mode is  $\sim 10$ msec  for a current sheet of  width $\sim 
 R_{NS}$  \citep{l03}.
Intermediate time scale $\sim 5$ msec observed in the GF may  be related 
to development of tearing mode in the
current sheets formed during onset of dynamical instability.

We associate the  overall duration of the spike $\sim 100 $ msec with the dynamical time
of the expanding magnetic cloud $\sim 2 \Gamma^2 c/R_{NS} \sim 25 $msec for $\Gamma \sim 20$. 
This is the  minimum  time it takes for expanding strongly magnetized bubble
to come into causal contact with itself. 
One expects that on this time scale magnetic cloud re-adjusts
its internal structure and relaxes to a minimal energy state.
This relaxation  occurs, \eg,  through reconnection, which in relativistic case
may proceed with the inflow velocity reaching the velocity of light  \citep{lu03}.

Finally,
typical time scale for the evolution of the tail of the GF, tens of seconds,
is well described by radiation leaking from  a plasma trapped  on closed magnetic field lines, TD95.

\subsection{ Quasi-thermal spectrum of the initial spike.}

Thermal spectrum results from 
 radiation escaping from  a (strongly magnetized) fireball that 
becomes optically transparent. Observer temperature of hundreds of keV
corresponds to rest frame temperature of $\sim 20 $ keV, when
plasma becomes optically thin  to pair production, boosted by 
Lorentz factor $\sim 10$ (\cite{g86,pac86},  see also \cite{nakar05}).

\subsection{Mass loading and terminal Lorentz factor}

In appendix \ref{warm} we consider dynamics of a hot, strongly magnetized
expanding flow carrying toroidal magnetic field. Flow is accelerated by magnetic and pressure forces, while both
matter inertia  and
 magnetic field energy density provide effective loading of the flow.
 In addition, in case of a large scale magnetic field considered here, there is 
 extra conserved quantity: magnetic flux. This plays an important role in the overall dynamics of 
 the flow \citep[\cf][]{KC84}.

If  source luminosity is $L$, mass loss rate is $\dot{M}_0$ and 
EMF is ${\cal E}$ (these are conserved quantities), then
the terminal Lorentz factor and terminal 
 magnetization parameter $\sigma_\infty$  are
\be  
\Gamma_\infty = { L \over \dot{M}_0 (1+ \sigma_\infty)},
, \,\,\,
\sigma_\infty = { {\cal E}^2 \over \Gamma_\infty \beta_\infty \dot{M}_0}.
\label{LK}
\ee
Which, formally, expresses the fact that magnetic field provides additional effective
loading (factor $1+ \sigma_\infty$), but the amount of loading 
depends non-trivially on  parameters of the flow.
In the 
 strongly relativistic limit, 
 outflow typically reaches \Alfven velocity (in fact fast magnetosonic), at which point
 Lorentz factor is related to the  terminal
  magnetization parameter as \citep{Michel71,gol69}
\be
 \Gamma_\infty = \sqrt{ \sigma_\infty},
 \label{Gammainf}
\ee

To estimate the maximum possible Lorentz factor, we note that the
minimum mass loading is determined by residual left-over pairs,
determined by equating annihilation and expansion rates \citep{g86,pac86,nakar05}
\be
M_{min} = m_e N_\pm = {4 \pi R_0 c t T_0^3 m_e \over \sigma _T T_\pm^3}=
2 \times 10^{17} {\rm g}
\label{M}
\ee
If most energy is in  magnetic form,
this corresponds to $\sigma_{max} = E / M_{min} c^2 = 4.5 \times 10^7$.
The corresponding 
maximum Lorentz factor is 
$
\Gamma_{max} = \sqrt{\sigma_{max}} =6.7 \times 10^3
$.
Lower limit on $\Gamma$  comes from the observed thermal temperature of the initial
spike:
\be
\Gamma_{min} \sim { T_{obs} \over T_\pm} \sim 10-20
\ee
 Thus, the  
flow must be  only weakly polluted by baryons.
This picture is in full agreement with TD01.
In what follows we adopt a minimum value of $\Gamma=10$ for numerical estimates. 
Then, estimating $E \sim L t_s$ and $M\sim \dot{M} t_s$ and using  Eq. (\ref{Gammainf}),
Eq. (\ref{Gammainf}) gives
\be
M \sim {E \over \Gamma \sigma c^2} =  {E \over \Gamma^3 c^2}\sim 10^{22} {\rm g}
\ee
This is the upper limit on amount of mass ejected during the GF.

\subsection{Plasma physics issues}

The proposed model of the GF builds on the models of Solar  CME and, similarly,
has a number of problematic  plasma physics issues, \citep[see, \eg][for review]{pf}
One is what is known as Aly-Sturrock paradox \citep{Aly,Sturrock}:
opening of  field lines, which is necessary to generate an outflow, requires an 
 increase in  the magnetic energy in the system, while the storage model of CMEs  requires magnetic 
 energy to decrease  during formation of magnetic cloud.
There is a number of ways  Aly-Sturrock paradox can be circumvented,   most important being 
magnetic reconnection which can change topology of the field line. 

Another problem is that injection of current occurs on a finite amount of magnetic flux 
so that 
in order to expand the newly formed magnetic cloud has to break through overlying  closed 
dipolar field
lines. This is achieved by reconnection at the null point at the leading edge of the magnetic cloud, \cf
''magnetic breakout'' model of \cite{antioch}. 
Reconnection transfers unshared magnetic flux associated with overlying dipolar
field to neighboring flux tubes, allowing the sheared filament to expand and erupt outward. 
The rate of reconnecting adjusts so that radial (as seen from the star)
propagation velocity is the \Alfven velocity, which in this case is nearly
the velocity of light.

\section{Radio afterglow: qualitative description}

\subsection{Expansion in SGR wind}

As magnetic cloud expands, it 
becomes transparent at $r_\pm$ and its pair density falls by many orders of magnitude. At this
point magnetic cloud
becomes strongly magnetically dominated. 
Initially magnetic cloud is topologically connected to the star, but eventually reconnection should happen
at the footpoints of the magnetic field lines, disconnecting the magnetic cloud  from the star.
At this point  magnetic cloud
d starts to expand into preexisting  SGR wind. 
It is expected that  SGR wind is strongly relativistic, with
Lorentz factors $\gg 10-20$. 
Thus, for intermediate baryon loading
(such that magnetic cloud expands with $\Gamma\sim 10-20$) 
magnetic cloud never overtakes the  wind, so that 
expansion
occurs as if in vacuum. 
Most of the magnetic energy is concentrated in a shell close to the bubble surface with thickness of the order of $c t_s\sim 10^9$ cm, where $t_s \sim 100 $ msec
is flare duration.

\subsection{Apparent constant non-relativistic
expansion velocity 
 is  due to relativistic strongly anisotropic expansion.}

Observations of constant expansion 
velocity from two to five weeks after the burst
have been interpreted as evidence in favor of large baryon loading, and, as 
a consequence, weak relativistic initial expansion  velocities \citep{Granot}.
A we argued above, 
this cannot be the case
due to compactness  constraints: flow must be relativistic with small  baryon loading.

Apparent  non-relativistic expansion velocity 
can be due to relativistic anisotropic expansion
 with little emission within the cone $1/\Gamma$
to the line of sight. If  emitting material is moving  relativistically
at an angle  $\theta \gg 1/\Gamma$, then apparent expanding velocity is 
$\beta _{app} = \beta \sin \theta / (1-\beta \cos \theta) \approx
 \beta \cot \theta/2$. To reproduce observed $\beta _{app} \sim 0.3-0.4 $,
 it required that outflow is directed away from the observer at an angle
 $\sim 135 ^{\circ}$.

We have arrived at a seeming contradictory picture: initially, during
the $0.2$ sec spike of the GF, expansion should be nearly isotropic, while
at later times, $\geq $ 1 week expansion is strongly anisotropic.
Thus, magnetic cloud should become strongly anisotropic
  between $0.2$ seconds and $7$ days. 
It is unlikely that anisotropy is achieved by internal magnetic stresses of the
freely expanding magnetic cloud: in relativistic case self-collimation is strongly suppressed
\citep[\eg][]{bogo01}.
Similarly, as we argue in \S \ref{spheromac}, collimating effects of the dipolar
magnetic field cannot account for strong anisotropy given the nearly spherically symmetric 
initial explosion.
We propose that expanding magnetic cloud becomes strongly anisotropic 
 due to fact that
magnetic fields of the cloud remain attached to the neutron star during most of the prompt phase.
Thus, at this times
the magnetic topology of the  expanding plasma is that
of a flux rope, see Fig. \ref{fluxrope}.

\subsection{Spheromac or flux rope?}
\label{spheromac}

Last decade two models were proposed for the structure of interplanetary
magnetic clouds ejected from the Sun: a magnetic flux rope and spheromac. 
The principal difference between the two is that in case of spheromac, 
magnetic cloud disconnects from the magnetic field of the Sun at early
stages of ejection and becomes quasi-spherical \citep[\eg][]{Low}, 
 while  flux rope remains attached to the Sun for a very long time
(even at the orbit of the Earth).
The two models lead to very different dynamics of the magnetic clouds. 
Presently,  spheromac model seems to be inconsistent with data  \citep[\eg][]{farrugia95}, while 
 magnetic flux rope model explains well 
 the internal magnetic 
structure ejected into interplanetary space \citep{Marubashi}.

\begin{figure}
\plotone{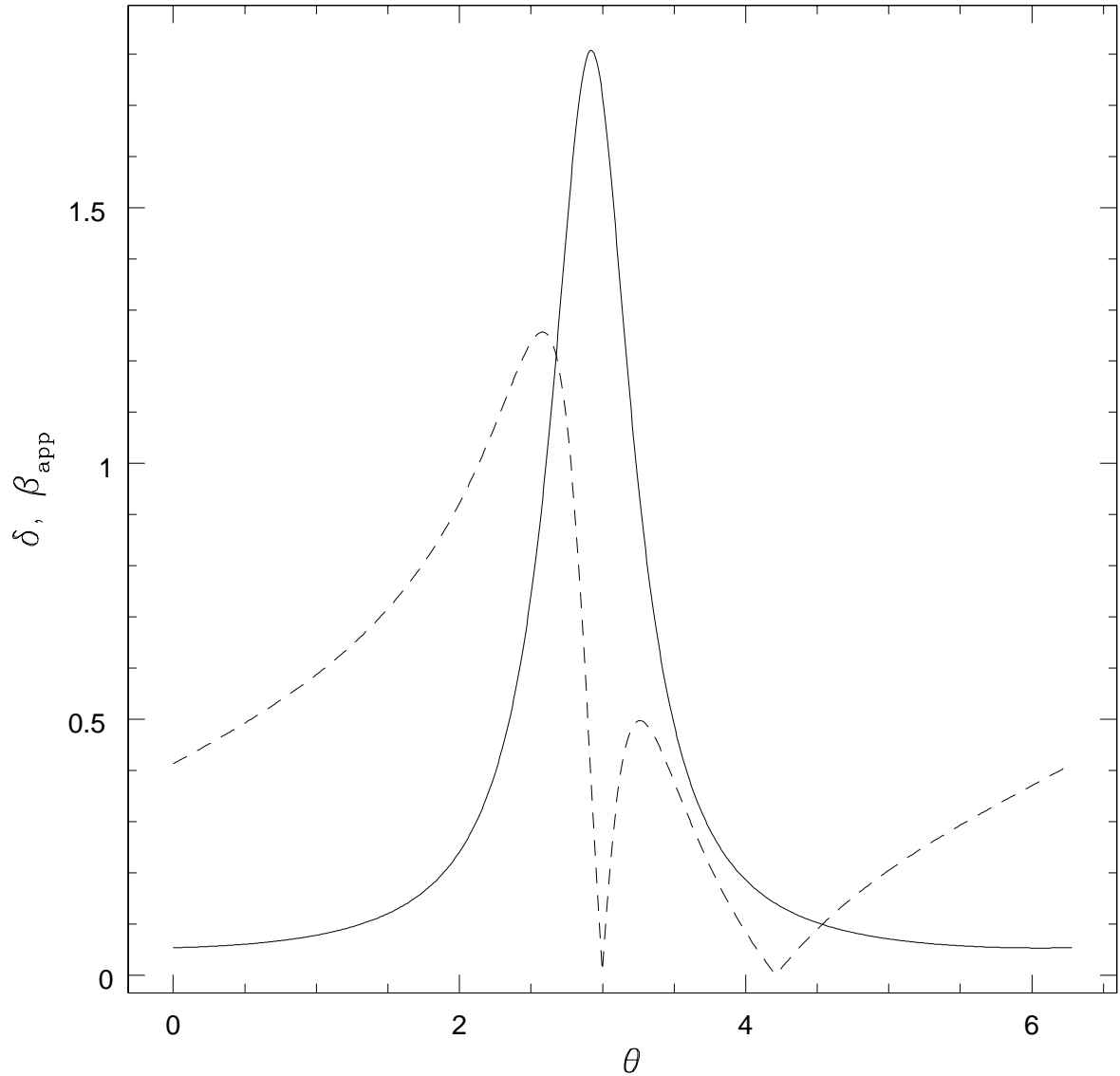}
\caption{
Doppler factor ${\cal D} =1/(\Gamma (1-\beta \cos \theta)$ (solid line) and apparent
transverse velocity (dashed line) for a relativistic  shock expanding with  Lorentz factor $\Gamma = 3$
and moving at $\theta_{ob} = 135 ^\circ$  with bulk Lorentz factor $\Gamma_{bulk}=2$ as a function of the
angle $\theta$ between the explosion direction  and the  emission point in the rest frame 
of the explosion for  emission points located in the plane containing  direction of  bulk motion and
line of sight. }
\label{betadelta}
\end{figure}

In  case of the GF, spheromac model seems to be 
inconsistent with data for the following reason. 
As we argued above, the initial explosion should be quasi-isotropic, while
at later times it should become strongly anisotropic and one-sided.
It unlikely that relativistic,  strongly  anisotropic explosions  are produced due to
collimating effects of internal magnetic field of the expanding
blob which is disconnected from the star:
for relativistic expansion 
collimation by internal magnetic field  
is kinematically suppressed \citep[\eg][]{bogo01},
and in any case cannot produce one-sided explosion.
Spheromac also cannot be efficiently  collimated by external dipolar magnetic field,
since inside spheromac
internal kinetic and magnetic pressures scale as 
$\propto B^2_{sph} \sim r^{-4}$ ($B_{sph} $ is a typical magnetic field inside 
spheromac), while $B_{dipolar}^2 \sim r^{-6}$. 
Thus, if the plasma to be ejected disconnects from the stellar magnetic field early on, 
one may expect only weak collimation and, as a result, 
weak relativistic bulk motion. Since the overall expansion should be
strongly relativistic, as we argued in \S \ref{relativistic},
observed emission will be   dominated by parts of the shock moving towards the observer, with somewhat
smaller Lorentz factor than average, but having large apparent velocity.
To illustrate the point, in Fig \ref{betadelta}  we plot 
Doppler factor $\delta  =1/(\Gamma (1-\beta \cos \theta)$ and apparent
transverse velocity for a relativistic  shock expanding with  Lorentz factor $\Gamma = 3$
and moving at $\theta_{ob} = 135 ^\circ$  with bulk Lorentz factor $\Gamma_{bulk}=2$ as a function of the
angle $\theta$ between the explosion direction  and emission point in the rest frame 
of the explosion and emission points located in the plane containing  direction of  bulk motion and
line of sight. Points on the shell with highest Doppler boosting 
have large apparent transverse velocities, in contradiction with observations.

On the other hand,  expanding magnetic cloud confined by a flux rope (with 
$B^2_{rope} \sim \sim r^{-4}$) may in principal provide collimation
(this would correspond to only regions near 
$\theta_{ob} \leq \pi$ in Fig. \ref{betadelta}   contributing to observed emission).
Details of this late  collimation need to be investigated numerically.
From observational point of view, at late times expansion should be confined to a fairly broad angle, of the order of $\pi$ radians, and not to thin, GRB-like
jet.

\section{Basic afterglow parameters}
\label{Basicafter}

\subsection{Geometry}

For numerical estimates
we chose $\theta_{ob} = 3 \pi/4 = 135^\circ$.
Then apparent velocity $\beta_{app} \sim \beta cot \theta/2 = 0.41 \beta$, Doppler
factor $\delta \sim 1/(2 \Gamma \sin^2 \theta/2) = 0.58/\Gamma$ and observer time is given
by $T=  2 r \sin^2 \theta/2/(\beta c) = 1.70 r/(\beta c)$.
We assume a strongly magnetized flow with total isotropic energy $E_{ej}\sim 10^{46}$ erg
which  reaches  terminal Lorentz factor $\Gamma =10$,
and is collimated into angle $d\Omega/(4 \pi) \sim .1$ (so that ''typical opening angle
is $\sim 36^\circ)$. We also 
normalize surrounding density to $n =1$ cm$^{_3}$.

\subsection{Typical radii and time scales}

Before the explosion, the magnetar is surrounded by a nearly empty bubble blown by the
magnetar wind with  a typical size
\be 
r_{s} \sim \left( { L_{w} \over 4 \pi n m_p c V_{NS}^2} \right)^{1/2}
\sim 1.2 \times 10^{16} {\rm cm}\, \left( {n \over 1 {\rm cm}^{-3}} \right) ^{-1/2}
\, \left( {V_{NS} \over 100 {\rm km s}^{-1}} \right)^{-1}
\ee
where $ L_{w} \sim 10^{34} $ erg/s is average spindown luminosity of SGR 1806 and 
$V_{NS} $  is  velocity of the neutron star.
Since magnetar wind is expected to  have Lorentz factor $\Gamma_w \gg 10$,
until $r_{s} $ magnetic cloud expands freely, without slowing down.
Note, that $r_{s} $ is the  minimum
distance from the neutron star to the shell, depending on the
relative orientation of the neutron star velocity and direction of explosion the time when magnetic cloud overcomes
the shell  can be larger by  factor of $\sim 2$.

Were it to expand in a constant density medium, the  
flow would starts to decelerate  at 
\be
r_{dec} \sim \left({3 E_{ej} \over d\Omega n m_p c^2 \Gamma_0^2} \right)^{1/3} \sim 
5.4 \times
10^{15} {\rm cm}  \left( {E \over 10^{46} {\rm erg s}^{-1}} \right)^{1/3}
\,  \left( {n \over 1 {\rm cm}^{-3}} \right) ^{-1/3} \,
\left( {\Gamma_0 \over 10} \right)^{-2/3}
\ee
Since $r_{dec} <  r_{s} $, magnetic cloud starts interacting with the a shell 
of shocked ISM plasma 
at $r\sim r_{s}$.
In observer time this occurs at
\be 
T\sim 1.7 {r_{s} \over c} = 8.2 {\rm days}
\ee

Since the amount of the mass contained in
a shell of shocked ISM plasma is $\sim n m_p r_{s} ^3$,
after encountering the shell Lorentz factor of the magnetic cloud falls down to
\be
\Gamma \sim \left( { 3 E  \over d\Omega n m_p c^2 r_{s}^3 } \right)^{1/2}
=2.8  \left( {E \over 10^{46} {\rm erg s}^{-1}} \right)^{1/2}\,
 \left( {V_{NS} \over 100 {\rm km s}^{-1}} \right)^{3/2}\,
 \left( {n \over 1 {\rm cm}^{-3}} \right) ^{1/4}\,
  \left( {d\Omega \over 0.1 \times 4\pi} \right)^{-1/2}
  \label{Gamma}
\ee
Doppler beaming factor at this point is $\delta = 0.2$.

Transition to the non-relativistic expansion occurs at 
\be
r_{nr} \sim \left({ 3 E_{ej} \over d\Omega n m_p c^2 } \right)^{1/3} \sim 1.7 \times 10^{16}
{\rm cm}
\ee corresponding to the observer time 
\be T_{nr} = 16.5 {\rm days}
\ee
$T_{nr}$ is an estimate of time when the velocity of the blast wave 
starts to deviate considerably from 
$c$.
It typically takes two times longer for the velocity to fall
below $0.5$ c, and for the Doppler factor to become within 15\% of unity. 
One expects to see a peak of emission at the moment when
  Doppler de-boosting becomes insignificant, approximately at $\sim 2 T_{nr}\sim 33 $ days.

\subsection{Total energy}
 Form the standard equipartition argument \citep{Pacholczyk},
  and taking into account Lorentz transformation
 of flux, $\propto \delta^{3-\alpha}$, and frequency,  $\propto \delta$,  the minimal energy of 
relativistic electrons plus magnetic field 
is 
\be
E_{min} =8 \times  10^{44} {\rm erg} 
\left( { \theta \over 65 ''} \right)^{9/7}  d_{15}^{17/7} \left( {F_\nu  \over 50 {\rm  mJy} } \right)^{4/7}
\left({\nu \over 8.5 {\rm GHz} } \right)^{2/7} \left({ \delta \over 0.2} \right)^{-2+4 \alpha/7}
\label{Emin}
\ee
where $\alpha \sim 0.5$ is the spectral index. 
This estimate is by  a factor $\delta^{-2+4 \alpha/7} \sim 15 $ 
larger than the one based on assumption of non-relativistic expansion.
Corresponding magnetic field (in laboratory frame) is 
\be
B_{min} =  0.1 \, {\rm G}\, \left( { \theta \over 65 ''} \right)^{-6/7} 
d_{15}^{-2/7} \left( {F_\nu  \over 50 {\rm  mJy} } \right)^{2/7}
\left({\nu \over 8.5 {\rm GHz} } \right)^{1/7} \left({ \delta \over 0.2} \right)^{-1-2 \alpha/7},
\ee
which is larger by  a factor $\delta^{-1-2 \alpha/7}\sim 6$
than the one based on assumption of non-relativistic expansion.
Note, that minimal energy argument address only energy in magnetic field and relativistic electrons.
It  is expected that most energy of the forward shock resides in protons, so that
the total energy in  the outflow may be  order(s) of magnitude larger than
(\ref{Emin}), bringing it in line with the total energy released in  $\gamma$-rays.

\section{Discussion}

In this paper we outlined a model of magnetar GF 
based on the analogy with Solar Coronal Mass Ejections.
Very short rise time scales of the GF indicate that
GF flare is driven by 
dissipation of energy stored in the magnetosphere right before the burst
(as suggested by \citep{l03}) and not in the crust of the neutron star
(as proposed by TD95). 
Initially, the  explosion is  loaded with pairs and is quasi-isotropic.
Magnetic field topology of expanding 
plasma resemble flux rope model of CMEs, which leads to late time collimation 
and  anisotropic
expansion.
Expanding magnetic cloud is strongly dominated by magnetic field, weakly baryon loaded, with $M \ll  E /c^2$, 
and strongly relativistic. 
Weeks after the flare,  magnetic cloud still
expands relativistically, strongly anisotropically and is moving away from the observer, 
resulting in apparent expansion
velocity $\beta_{app} 
\sim 0.4$. At approximately $33$ days, corresponding  to a  bump in the light curve,
expansion  velocity falls below $c/2$.
The main prediction of the model is that we may see a medium-energy flare with a very bright
radio afterglow, when the explosion will be beamed towards the Earth. 
Statistically, it should take approximately 
$\sim 1/d\Omega
\sim 10$ GFs. 

Magnetospheric dissipation following  plastic deformation of the crust
is also consistent with suggestion of \cite{Jones02}  that that neutron-star matter cannot exhibit
brittle fracture and should instead experience only plastic deformations. 
On the other hand, complexity of earthquakes, especially so called deep
focus  earthquakes occurring at high pressures, indicates that \cite{Jones02}
argument is not the end of the story. For example, crust response may depend
on value of a strain, being plastic at low strains and brittle at high
strains \citep{Frohlich}. 

Can crustal fractures model be consistent with above arguments?
One possible way is to invoke small scale, $\sim $ 100 meters,
initial crustal deformation (so that the rise time of the GF is short enough),
which triggers larger scale deformations in an avalanche-type process
(TD95, TD01).  In addition,  relatively bright and
long-lived afterglow following  August 27 flare is well fitted by
 the  deep crustal heating model \citep{lyubar}.
The crustal fracture model also has a better chance of explaining
post-flare activity ( August 27 and  March 5th  flares  followed  by
a burst-active period) in analogy
with earthquake aftershocks.

One possible way to distinguish the models is that
reconnection-type  events may
be accompanied by  coherent  radio emission  resembling
 solar type-III radio bursts \cite{lyu02}.
 The radio emission should have  correlated pulse profiles
 with X-rays, narrow-band-type radio spectrum with
 $\Delta \nu \leq \nu$  with the typical frequency $ \nu \geq 1 $ GHz,
 and a drifting central frequency. 
 This requires catching a burst in simultaneous radio and X-ray
 observations. 

\begin{acknowledgements}
We would like to thank Roger Blandford, Yuri Lyubarky and  Christofer Thompson for numerous discussions.
\end{acknowledgements}

\appendix

\section{Dynamics of magnetized pair-loaded  flows }
\label{warm}

In this  section we consider  dynamics of a hot relativistic,
spherically symmetric,  stationary outflow carrying toroidal magnetic field.
 Since the energy release during GF lasts $\sim 0.1$ s, 
on small time scales and for radii $\leq 10^9$ cm flow may be considered stationary. 
In addition, internal energy density in the expanding magnetic cloud scales slower with the radius, 
$\propto r^{-4}$,
than energy density of the dipolar field $\propto r^{-6}$, so that dynamical effects
of dipolar magnetic field quickly become negligible. At this stage expansion is quasi-isotropic,
as discussed in \S \ref{Initial}. Rotation and dynamical effects of the poloidal
magnetic field are  neglected.

The  dynamics of such  a warm magnetized wind
  is    controlled by three parameters:
energy $L$, 
 mass flux $\dot{M} $  and the electro-motive force ${\cal E}$ produced by the expanding magnetic cloud.
 Total energy flux may be divided in two 
forms: mechanical $L_M$ and  electromagnetic $L_{EM}$ luminosities.
\be
L = L_M + L_{EM}
\ee
We wish to understand how the parameters of a fully relativistic
flow   (velocity $\beta$, 
pressure $p$,
magnetization $\sigma$) evolve for an arbitrary ratio of both $L_{EM} / L_M$ 
and $p/\rho$ ($\rho$ is the rest-frame mass density).

The 
asymptotic evolution of the flow
is determined by  conserved quantities which 
may be chosen as the  total luminosity $L$,  the mass flux $\dot{M}$
and the  EMF  ${\cal E}$.
 Thus, the central source
works both as thruster and as a dynamo.

The formal treatment of the problem
 starts with the 
set of relativistic  MHD  equations which 
can be written in terms of conservation laws. In coordinate form and
assuming a stationary, radial,
 spherically symmetric outflow with toroidal magnetic field, relativistic  MHD  equations give
\ba 
&&
{1\over r^2} \partial_r \left[ r^2 ( w+b^2) \beta \Gamma^2 \right] =0
\label{x4} \\ &&
{1\over r^2} \partial_r 
\left[r^2\left((w+b^2) \beta^2 \Gamma^2 + (p+b^2/2)\right) \right] - 
{2 p \over r}=0
\label{x5} \\ &&
{1\over r} \partial_r  \left[r b \beta \Gamma \right] =0
\label{x51} \\ &&
{1\over r^2 } \partial_r  \left[r^2 \rho  \beta \Gamma \right] =0
\label{x6}
\ea
The above relations can be simplified  by defining
\be 
L = 4 \pi r^2 \beta \,{\Gamma }^2\,\left( b^2 + {\Gamma_{a} \over \Gamma_{a} -1} 
 \,p + \rho  \right) , \,\,\,
\dot{M} =  4 \pi r^2 \beta \,\Gamma \,\rho  , \,\,\,
{\cal E}= 2 \sqrt{\pi} r\, \beta \,\Gamma \, b
\ee
where  we assume that
fluid is polytropic with adiabatic  index $\Gamma_{a}$: $w =\rho+ {\Gamma_{a} \over \Gamma_{a} -1} p$,
 ${\cal E}$ is the electromotive force.

It is convenient to introduce two other parameters: the 
magnetization parameter $\sigma$ as the ratio of the rest-frame
magnetic and particle energy-density
 and a fast magnetosonic  wave  phase velocity
$\beta_f$ 
\be 
\sigma={ b^2 \over w} = {{\cal E}^2  \over L \beta - {\cal E}^2} , \,\,\,
\beta_f^2=  
{\sigma \over 1+ \sigma} + {\Gamma_{a} p \over (1+ \sigma) w}=
\left( \Gamma_{a}-1 \right) \left( 1 - { \Gamma \dot{M} \over L}  \right)+
(2-\Gamma_{a}) { {\cal E}^2  \over L \beta}
\label{pp}
\ee
Using the  three conserved quantities $L, \, \dot{M}$ and $ {\cal E}$
the evolution equation becomes
\be
{ 1 \over 2  \beta^2 \Gamma  }
\partial_ {r }  \Gamma= { (\Gamma_{a}-1)
\left( \beta L - \beta \Gamma \dot{M}  - {\cal E}^2 \right) \over
 r\, \left(\beta L  ( \beta^2 +1 -\Gamma_{a}) + (\Gamma_{a}-1) \beta \Gamma  \dot{M}  
-(2-\Gamma_{a}) {\cal E}^2 \right) }
\label{gt}
\ee
Eliminating 
${\cal E}$ in favor of $\Gamma_f$ we 
get a particularly transparent form for the evolution of Lorentz
factor
\be
{  \left( {\Gamma }^2 - {{{\Gamma }_f}}^2 \right) \over \beta^2  \Gamma^3}
\partial_r \Gamma=
{ 2  p \Gamma_{a} \over ( w - \Gamma_{a} p) r } 
\label{dg}
\ee
Equation (\ref{dg}) is   nozzle-type   flow
(e.g. \citep{LLIV}).
The lhs of  eq. (\ref{dg}) contains a familiar 
critical  point at the sonic transition $\Gamma= \Gamma_f$.
The positively defined  rhs  describes the  evolution
of Lorentz factors due to kinetic pressure effects. 
In the case of purely radial expansion the magnetic gradient
forces are exactly balanced by the hoop stresses, so that
magnetic field does not contribute to acceleration.
From Eq. (\ref{dg}) it follows that 
super-fast-magnetosonic flows accelerate while
sub-fast-magnetosonic flows
 decelerate. 
It also  follows that 
terminal Lorentz factor of the flow  is determined by  the condition
$\partial_r \Gamma =0$ which implies  that either 
$p=0$ or $\beta_{\infty}=0$.
Condition $\beta_{\infty} =0$ can be reached only for subsonic flows with ${\cal E}=0$.
Neglecting the  $\beta_{\infty} =0$ solution, the terminal
velocity of magnetized flow is determined  by the condition:
\be
L=\Gamma_{\infty}  \dot{M} + { {\cal E}^2 \over \beta _{\infty}}
\label{L1}
\ee
For each set of parameters ($L$, $ \dot{M}$ and ${\cal E}$) there
are  generally two solutions of Eq. (\ref{L1}) for the terminal four-velocity. The only 
exception is the 
case of zero magnetization,  $ {\cal E}= 0$,
when the terminal (supersonic) Lorentz
factor is uniquely  determined by $ \Gamma_{\infty} = L/ \dot{M}$.

For non-zero magnetization the terminal 
velocity cannot be determined uniquely from  given $L$, $\dot{M}$ and ${\cal E}$.
Solutions exist only for
$ L  / \dot{M}$ larger than some critical value
 $L / \dot{M}=(1 - ( {\cal E}^2/L)^{2/3})^{-3/2}$,
corresponding to  $\beta_\infty =  ( {\cal E}^2/L)^{1/3}$.
For fixed $\dot{M}$  and ${\cal E}$ the  minimum energy  loss is reached
at $\beta_{min}$.
 Assumption  $\beta_\infty = \beta_{min}$, 
then gives
\be
\left( 1 + ( \beta_\infty \Gamma_\infty \sigma_\infty)^{2/3} \right)^{3/2} = 
{ \Gamma_\infty \sigma_\infty \over \beta_\infty^2}
\label{W}
\ee
which in the strongly relativistic limit gives
Michel solution
\be
 \Gamma_\infty = \sqrt{\sigma_\infty}.
 \ee

We  can also   relate
the terminal magnetization $\sigma_\infty$ to the magnetization at the
source -  more specifically to magnetization at the sonic point
$\sigma_f$:
\be
\sigma_f =  
\left\{
\begin{array}{lll}
{\sigma_\infty \over \Gamma_{a}-1 } & = 3 \sigma_\infty &
\mbox{ if  $ \sigma_\infty \ll 1$} \\
{ (4- \Gamma_{a}) \sigma_\infty \over 2} & = {4\over 3} \sigma_\infty & 
\mbox{ if $ \sigma_\infty \gg  1$}
\end{array} \right.
\ee
Thus, we always have $\sigma_\infty < \sigma_f$, but they
remain of the same order of magnitude:
the magnetization of the flow changes only slightly as the
flow propagates away from the launching point to infinity. 
The reason for constant $\sigma$ in the supersonic regime is that
both in the case $p \gg \rho $ (linear acceleration stage) and $p \ll \rho $
(coasting stage)
 the plasma and the magnetic field
energy densities in the  flow change with the same radial dependence ($\sim r^{-4}$
and $\sim r^{-2}$ correspondingly).

\begin{figure}
\plottwo{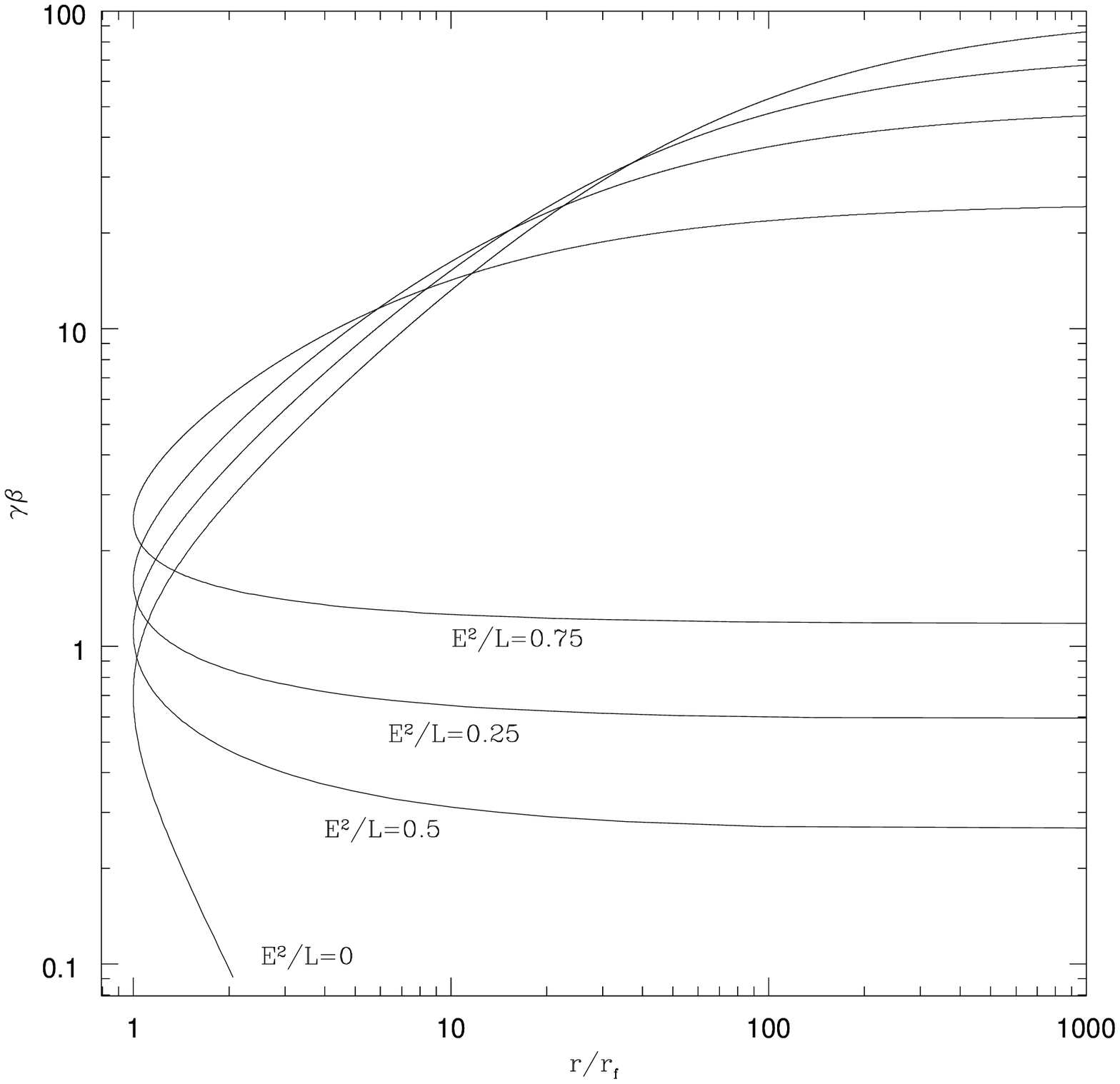}{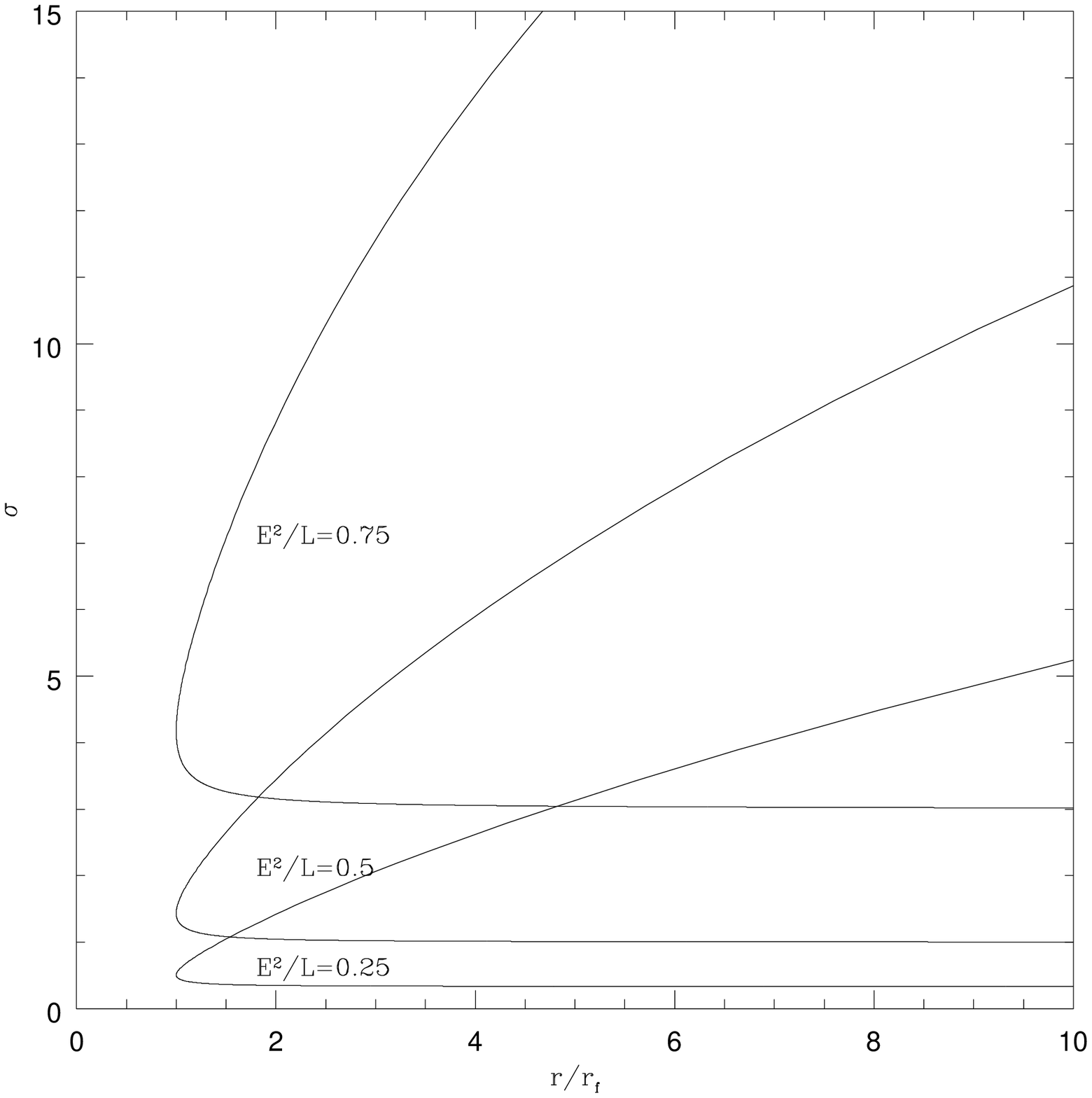}
\caption{  Evolution of strongly magnetized and  hot   flows.
(a) 
Four-velocities 
 of flows
 are given 
as functions of  $r/r_f$ for $L/\dot{M}=100$ and 
 different values of the parameter ${ {\cal E}^2 \over L}$.
 Flows start at $ r= r_f$ with $\beta = \beta_f$; supersonic
flows first  accelerate as $ \beta \gamma \sim r$, reaching a 
terminal value given by the larger root of  eq. (\ref{L1}), 
while subsonic decelerate initially as $\beta \sim r^{-2}$  
reaching asymptotic value given by the smaller root of eq. (\ref{L1}).
(b) Magnetization parameter $\sigma$.
For supersonic flows (lower branch)
 the magnetization remains constant, reaching
$\sigma_\infty = \left( 1 - { {\cal E}^2 \over L} \right)^{-1} $
as   $r \rightarrow \infty $.
Subsonic flows become strongly magnetized as they expand (upper branch); the
  magnetization parameter increases  $\sigma \sim r^{2/3}$.
}
\label{rofu1}
\end{figure}

For arbitrary flow parameters the evolution equations are integrated numerically
(Fig. \ref{rofu1}).
Given the evolution of the flow and the relation for local $\sigma$ 
 we can find the 
evolution of the magnetization parameter (Fig. \ref{rofu1}.b).

\newpage

\end{document}